\documentclass[conference]{IEEEtran}

\IEEEoverridecommandlockouts

\usepackage{cite}
\usepackage{orcidlink}

\usepackage{xspace}

\usepackage{csquotes}
\usepackage{enumitem}

\usepackage{graphicx}
\usepackage{pgfplots}
\usepackage{tikz}

\usepackage{amsmath}
\usepackage{amssymb}
\usepackage{amsthm}
\usepackage{mathtools}
\usepackage{bm}
\usepackage{mathrsfs}
\usepackage{mleftright}
\usepackage{siunitx}
\usepackage[only,llbracket,rrbracket]{stmaryrd}

\usepackage{hyperref} %

\usepackage{glossaries} %
\glsdisablehyper

\makeatletter
\newcommand*{\itodo}[1]{\tikzexternaldisable\@todo[inline]{#1}\tikzexternalenable}
\newcommand*{\bigtodo}[2]{\tikzexternaldisable\@todo[inline, caption={#1}]{\begin{minipage}{\linewidth-4pt}\textbf{#1:}\\#2\end{minipage}}\tikzexternalenable}
\newcommand*{\itemtodo}[1]{\tikzexternaldisable\@todo[inline]{\unexpanded{\unexpanded{\begin{minipage}{\linewidth}\begin{itemize}#1\end{itemize}\end{minipage}}}}\tikzexternalenable}
\newcommand*{\bigitemtodo}[2]{\tikzexternaldisable\@todo[inline, caption={#1}]{\unexpanded{\unexpanded{\begin{minipage}{\linewidth}\textbf{#1:}\begin{itemize}#2\end{itemize}\end{minipage}}}}\tikzexternalenable}
\makeatother

\pgfplotsset{compat=1.18}
\usepgfplotslibrary{units}
\usetikzlibrary{external,pgfplots.units,arrows.meta,matrix,chains,positioning,intersections,calc,fit,dsp}
\pgfplotsset{table/search path={results/},}
\definecolor{wongblue}{RGB}{0, 114, 178}
\definecolor{wongorange}{RGB}{230, 159, 0}
\definecolor{wonggreen}{RGB}{0, 158, 115}
\definecolor{wongpurple}{RGB}{204, 121, 167}
\definecolor{wonglightblue}{RGB}{86, 180, 233}
\definecolor{wongvermillion}{RGB}{213, 94, 0}
\definecolor{wongyellow}{RGB}{240, 228, 66}

\definecolor{matlabblue}{rgb}{     0,     0.447, 0.741}
\definecolor{matlaborange}{rgb}{   0.85,  0.325, 0.098}
\definecolor{matlabyellow}{rgb}{   0.929, 0.694, 0.125}
\definecolor{matlabpurple}{rgb}{   0.494, 0.184, 0.556}
\definecolor{matlabgreen}{rgb}{    0.466, 0.674, 0.188}
\definecolor{matlablightblue}{rgb}{0.301, 0.745, 0.933}
\definecolor{matlabred}{rgb}{      0.635, 0.078, 0.184}

\pgfplotscreateplotcyclelist{wongcolors}{{wongblue}, {wongorange}, {wonggreen}, {wongpurple}, {wonglightblue}, {wongvermillion}, {wongyellow}}

\dspdeclareoperator{dspshapexor}{
	\pgfmoveto{\pgfpointadd{\centerpoint}{\pgfpoint{0pt}{-\radius}}}
	\pgflineto{\pgfpointadd{\centerpoint}{\pgfpoint{0pt}{ \radius}}}

	\pgfmoveto{\pgfpointadd{\centerpoint}{\pgfpoint{-\radius}{0pt}}}
	\pgflineto{\pgfpointadd{\centerpoint}{\pgfpoint{ \radius}{0pt}}}

	\pgfusepathqstroke
}

\tikzset{%
	block/.style     = {draw,rectangle,align=center,inner sep=2mm},
	bigblock/.style  = {draw,rectangle,align=center,inner sep=2mm,minimum height=2.5em},
	hiergroup/.style = {draw,line width=0.3pt,inner sep=5mm,rectangle,rounded corners},
	dspxor/.style    = {shape=dspshapexor,line cap=rect,line join=rect,line width=\dspblocklinewidth,minimum size=\dspoperatordiameter},
}

\pgfplotsset{
	discard if not/.style 2 args={
		x filter/.code={
			\edef\tempa{\thisrow{#1}}
			\edef\tempb{#2}
			\ifx\tempa\tempb
			\else
				
			\fi
		}
	}
}

\makeatletter
\newcommand{\@pltref}[1]{\tikzexternaldisable\ref{#1}\tikzexternalenable}
\newcommand{\@@pltref}[1]{(\tikzexternaldisable\ref{#1}\tikzexternalenable)}

\newcommand{\pltref}{\@ifstar\@pltref\@@pltref}
\makeatother

\theoremstyle{plain}

\theoremstyle{definition}

\theoremstyle{remark}

\renewcommand*{\vec}[1]{\bm{#1}}
\newcommand*{\matr}[1]{\bm{#1}}
\newcommand{\bmat}[1]{\ensuremath{\begin{bmatrix}#1\end{bmatrix}}}

\newcommand{\polarcode}{\operatorname{\mathcal{PC}}}

\DeclareMathOperator*{\argmax}{arg\,max}

\DeclareMathOperator{\E}{\mathbb E}
\DeclareMathOperator{\He}{\mathbb H} %
\DeclareMathOperator{\MI}{\mathbb I}

\DeclarePairedDelimiter\abs{\lvert}{\rvert}
\DeclarePairedDelimiter\idxset{\llbracket}{\rrbracket}

\DeclarePairedDelimiter\set{\{}{\}}

\DeclarePairedDelimiterX{\infdivx}[2]{(}{)}{#1\;\delimsize\|\;#2}

\DeclareSIUnit{\belc}{Bc}
\DeclareSIUnit{\belm}{Bm}
\DeclareSIUnit{\bit}{bit}
\DeclareSIUnit{\sample}{S}
\DeclareSIUnit{\bpcu}{bpcu}

\newacronym{ASK}{ASK}{amplitude-shift keying}
\newacronym{AWGN}{AWGN}{additive white Gaussian noise}
\newacronym{BER}{BER}{bit error rate}
\newacronym{BEC}{BEC}{binary erasure channel}
\newacronym{BICM}{BICM}{bit-interleaved coded modulation}
\newacronym{biDMC}{biDMC}{binary-input discrete memoryless channel}
\newacronym{bpcu}{bpcu}{bits per channel use}
\newacronym{BPSK}{BPSK}{binary phase-shift keying}
\newacronym{CCDM}{CCDM}{constant composition distribution matching}
\newacronym{CRC}{CRC}{cyclic redundancy check}
\newacronym{diMC}{diMC}{discrete-input memoryless channel}
\newacronym{DMC}{DMC}{discrete memoryless channel}
\newacronym{DM}{DM}{distribution matching}
\newacronym{DPC}{DPC}{dirty paper coding}
\newacronym{dSNR}{dSNR}{design signal-to-noise ratio}
\newacronym{FEC}{FEC}{forward error control}
\newacronym{FER}{FER}{frame error rate}
\newacronym{HY}{HY}{Honda-Yamamoto}
\newacronym{iff}{iff}{if and only if}
\newacronym{iid}{i.i.d.}{independent and identically distributed}
\newacronym{IM}{IM}{intensity modulation}
\newacronym{LDPC}{LDPC}{low-density parity-check}
\newacronym{LLPS}{LLPS}{linear layered probabilistic shaping}
\newacronym{LLR}{LLR}{log-likelihood ratio}
\newacronym{MC}{MC}{Monte Carlo}
\newacronym{MI}{MI}{mutual information}
\newacronym{MLC}{MLC}{multilevel coding}
\newacronym{MLHY}{MLHY}{multilevel Honda-Yamamoto}
\newacronym{MLPC}{MLPC}{multilevel polar coding}
\newacronym{MMSE}{MMSE}{minimum mean square error}
\newacronym{MSD}{MSD}{multistage decoding}
\newacronym{OOK}{OOK}{on-off keying}
\newacronym{PAM}{PAM}{pulse-amplitude modulation}
\newacronym{PAS}{PAS}{probabilistic amplitude shaping}
\newacronym{PS}{PS}{probabilistic shaping}
\newacronym{PCPAS}{PC-PAS}{polar-coded probabilistic amplitude shaping}
\newacronym{QAM}{QAM}{quadrature-amplitude modulation}
\newacronym{RCUB}{RCUB}{random coding union bound}
\newacronym{SCL}{SCL}{successive cancellation list}
\newacronym{SC}{SC}{successive cancellation}
\newacronym{SE}{SE}{spectral efficiency}
\newacronym{SIR}{SIR}{signal-to-interference ratio}
\newacronym{SMI}{SMI}{symmetric mutual information}
\newacronym{SNR}{SNR}{signal-to-noise ratio}
\newacronym{TCMPAS}{TCM-PAS}{trellis-coded modulation probabilistic amplitude shaping}
\newacronym{wlog}{w.l.o.g.}{without loss of generality}

\newcommand{\eg}{e.g.}
\newcommand{\ie}{i.e.}

\newcommand*{\Huuy}{\He(U_i|\vec U_\bbi, \vec Y)}
\newcommand*{\Huu}{\He(U_i|\vec U_\bbi)}

\newcommand*{\bbN}{{\idxset{N}}}

\newcommand*{\bbi}{{\idxset{i-1}}}

\makeatletter
\input{tikz/fer_ask8.dpth}
\input{tikz/fer_pam4.dpth}
\makeatother

\begin{document}

\title{Improved List Decoding for Polar-Coded Probabilistic Shaping}

\tikzexternaldisable
\author{\IEEEauthorblockN{Constantin Runge\,\orcidlink{0000-0001-8324-3945},
Thomas Wiegart\,\orcidlink{0000-0002-8498-6035},
Diego Lentner\,\orcidlink{0000-0001-6551-8925}}
\IEEEauthorblockA{Institute for Communications Engineering,
Technical University of Munich,
80333 Munich, Germany \\
\{constantin.runge, thomas.wiegart, diego.lentner\}@tum.de}
\thanks{This work was supported in part by the German Federal Ministry of Education and Research (BMBF) under the Grant 6G-life, identification number 16KISK002, and by the German Research Foundation (DFG) under Projects 390777439 and 509917421.}
}

\maketitle
\tikzexternalenable

\begin{abstract}
A modified \gls{SCL} decoder is proposed for polar-coded probabilistic shaping.
The decoder exploits the deterministic encoding rule for shaping bits to rule out candidate code words that the encoder would not generate.
This provides error detection and decreases error rates compared to standard \gls{SCL} decoding while at the same time reducing the length of the outer cyclic redundancy check code.
\end{abstract}
\glsresetall


\section{Introduction}

Reliable and power-efficient communication over noisy channels requires the code words to approximate the capacity-achieving input distribution.
This can be achieved by \gls{PS} which can be implemented via, e.g., many-to-one mappings \cite[Sec.~6.2]{Gallager68}, trellis shaping \cite{forney92}, \gls{PAS}~\cite[Sec.~II]{Bocherer15,Gultekin20}, and others.
A recent scheme is \gls{MLHY} \gls{PS} \cite{Iscan18comml,Runge22} based on polar codes \cite{Stolte02, Arikan09} and their extension to asymmetric channels \cite{Honda13} via nested polar codes \cite{Korada10b}.

\Gls{PAS} combines systematic \gls{FEC} codes with \gls{DM} algorithms to approach capacity with flexible rate adaptation.
It requires that the target distribution $P_X$ factors as $P_A \cdot P_S$ where $P_S$ is the uniform binary distribution.
We use ``$A$'' and ``$S$'' to refer to the amplitude and sign of $X$, respectively, but more general choices are permitted.
For instance, \gls{CCDM} \cite{Schulte16} encodes the information bits into a string $\vec A$ of amplitudes  distributed according to $P_A$.
The uniformly distributed parity bits of the \gls{FEC} code are mapped to signs $\pm1$ that multiply the amplitudes.
\Gls{PAS} does not allow for asymmetric distributions $P_X$ in general.\footnote{One can choose the code to perform joint shaping and coding on $S$ but it then seems more natural to do this directly rather than via \gls{PAS}.}

Polar codes are linear block codes that can achieve capacity over symmetric \glspl{DMC} and that can be constructed, encoded, and decoded efficiently.
However, the reliability scales poorly in the block length as compared to low-density parity-check (LDPC) codes.
Remarkably, simple modifications based on concatenating a \gls{CRC} outer code and applying  list decoding \cite{Tal15} make polar codes competitive at short to moderate block lengths. We focus on binary polar codes.

The polarization-based coset coding scheme of \cite{Korada10b,Honda13}, called \gls{HY} coding, performs joint \gls{DM} and \gls{FEC} to
generate code words distributed according to general $P_X$.
The method achieves capacity over general \glspl{biDMC} \cite{Honda13}, and also over general \glspl{DMC} when combined with \gls{MLC} \cite{Imai77,Seidl13,Runge22} or non-binary kernels \cite{Sasoglu09,Park13}.
\Gls{HY} coding can be adapted to channels and channel inputs with memory \cite{Wang15,Shuval19,Sasoglu19}.
At short to moderate block lengths, \gls{HY}-based coding shows excellent performance~\cite{Wiegart19,Runge22}.

The motivation for this paper is as follows. Observe that \gls{DM} induces a concatenated code structure, e.g., \gls{CCDM} ensures that all code words have the same empirical distribution or type. This  allows a \gls{SCL} decoder to exclude candidate code words with the wrong type \cite{Prinz17}. The paper \cite{Runge22} uses this idea to show that \gls{PCPAS} as in \cite{Prinz17} can sometimes outperform \gls{MLHY} coding with a shorter \gls{CRC} code.
We here apply this idea to \gls{HY} coding and propose two methods to exclude invalid code words at the decoder.

This paper is organized as follows.
Sec.~\ref{sec:pre} reviews notation and polar coding concepts.
Sec.~\ref{sec:hy} provides a more detailed look at \gls{HY} codes and their en-/decoding.
Sec.~\ref{sec:main} describes our proposed modifications and discusses complexity.
Finally, Sec.~\ref{sec:results}  compares \gls{FER} performances of standard decoding and improved decoding. Sec.~\ref{sec:results} further contains a statistical analysis of the additional complexity induced by our modifications.

\section{Preliminaries}
\label{sec:pre}

Random variables are written with upper case letters such as $X$.
Their alphabet, distribution, and realizations are written as $\mathcal{X}$, $P_X$, and $x$, respectively.
The $N$-fold product distribution is denoted as $P_X^N$.
Vectors are denoted by bold symbols such as $\vec x$.
$\abs{\mathcal X}$ is the cardinality of $\mathcal X$.
An index set from $1$ to $N$ is denoted as $\bbN \triangleq \set{1,\dots,N}$.
A set $\mathcal S$ may select entries of a vector, creating a substring $\vec x_{\mathcal S}$ with length $\abs{\mathcal S}$, \eg, $\vec x_\bbN$.
We denote by $\He(X)$, $\He(X|Y)$, and $\MI(X; Y)$ the entropy of $X$, the entropy of $X$ conditioned on $Y$, and the \gls{MI} between $X$ and $Y$, respectively.
A polar code with set of frozen indices $\mathcal F$ and block length clear from the context is denoted as $\polarcode(\mathcal F)$.

\subsection{Polar Codes}
\subsubsection{Polarization}

Polar codes are linear block codes of length $N=2^n$ and dimension $K$.
They are defined via the self-inverse polar transform $G_N$ that maps a vector $\vec u\in\mathbb F_2^N$ to a code word $\vec x\in\mathbb F_2^N$,
\begin{equation}
	\vec x = \vec u \matr G_N \textnormal{ with } \matr G_N = \matr B_N \bmat{1 & 0\\ 1 & 1}^{\otimes n}
\end{equation}
where $\matr B_N$ is the bit-reversal matrix as in \cite{Arikan09}, and where $\matr F^{\otimes n}$ is the $n$-fold Kronecker product of $\matr F$.
The code word $\vec x$ is transmitted over $N$ uses of a channel $P_{Y|X}$ resulting in a vector of channel observations $\vec y\in\mathcal Y^N$.

With $\mathcal U(\mathcal X)$ being the uniform distribution on $\mathcal X$, assume $P_{XY} = \mathcal U(\mathbb F_2) P_{Y|X}$ with a \gls{biDMC} $P_{Y|X}$ and consider the index sets
\begin{align}
	\mathcal H_{U|Y} &= \set{i\in\bbN: \Huuy > 1 - \delta} \\
	\mathcal L_{U|Y} &= \set{i\in\bbN: \Huuy < \delta}
\end{align}
with $0 < \delta < 1$.
By construction of $\matr G_N$, these sets polarize to \cite{Arikan09}
\begin{align}
	\lim_{N\to\infty} \frac1N \abs{\mathcal H_{U|Y}} &= \He(X|Y) \\
	\lim_{N\to\infty} \frac1N \abs{\mathcal L_{U|Y}} &= 1 - \He(X|Y) \textnormal{.} \label{eq:polar_symmetric_capacity}
\end{align}
Note that this holds for general $P_{XY}$ regardless of the physical interpretation.

This motivates the following coding scheme.
The positions of the vector $\vec u$ are partitioned into index sets $\mathcal F$ and $\mathcal R$ which are inspired by $\mathcal H_{U|Y}$ and $\mathcal L_{U|Y}$, respectively.
$\mathcal R$ holds the $K$ positions with lowest $\Huuy$, \ie, the most reliable bits, and $\mathcal F$ holds the remaining $N-K$ so-called frozen positions.
The bits of $\mathcal F$ are fixed to $\vec u_{\mathcal F} = 0$ and the bits of $\mathcal R$ hold the data, $\vec u_{\mathcal R} = \vec d$.
By construction, $\Huuy\approx0$ for $i\in\mathcal R$.
As $u_i=0$ is known at the receiver for $i\not\in\mathcal R$, for each $i\in\bbN$ the previous bits $\vec u_\bbi$ are either known or can be reliably decoded.
A so-called \gls{SC} decoder computes $P_{U_i|\vec U_\bbi, \vec Y}$, $i\in\bbN$, one after another from $P_{X|Y}^N(\cdot|\vec y)$ and decides at each step
\begin{equation}
    \label{eq:sc_decoding}
    \hat u_i = \begin{cases} 0 \textnormal{, } & i \in \mathcal F \\ \argmax_{u} P_{U_i|\vec U_\bbi, \vec Y}(u|\vec{\hat u}_\bbi, \vec y) \textnormal{, } & i \in \mathcal R \,.\end{cases}
\end{equation}

Polar coding extends straightforwardly to multilevel codes \cite{Seidl13}.
By exploiting conceptual similarities of \gls{MSD} and \gls{SC} decoding, one can analyze the resulting code as one polar code.

\subsubsection{Successive Cancellation List Decoding}

The special structure of \gls{SC} decoding for polar codes allows for an efficient extension to list decoding \cite{Tal15}.
The \gls{SCL} decoder successively computes $P_{U_i|\vec U_\bbi, \vec Y}$ identically to the \gls{SC} decoder.
Instead of deciding for the most likely $\hat u_i$ every time, a list of possible partial candidates $\vec{\hat u}_\bbi^{(\ell)}$, $\ell\in\idxset{L}$, with list size $L$ is maintained.
At each information position $i\in\mathcal R$, the decoder extends every partial candidate $\vec{\hat u}_\bbi^{(\ell)}$ with both $\hat u_i=0$ and $\hat u_i=1$ and collects the resulting $\vec{\hat u}_{\idxset{i}}$ in the list.
These candidates are each assigned a path metric which tracks their respective likelihood.
When the number of partial candidates exceeds $L$, only the $L$ most likely paths are kept.

After the decoder visited all positions $i\in\bbN$ it outputs a list of $L$ code word candidates of which the most likely candidate is chosen as the final estimate.
This can be combined with an additional outer \gls{CRC} error detection code.
Then, the decoder chooses as final estimate the most likely candidate which satisfies the \gls{CRC} parity checks.
With a proper choice of \gls{CRC} code, the error correction performance of polar codes under \gls{SCL} decoding can be significantly improved \cite{Tal15}.

\section{Polar-coded Probabilistic Shaping}
\label{sec:hy}

The code construction can be modified to generate channel inputs $\vec x$ with non-uniform distribution $P_X $\cite{Honda13}.
For this, one considers polarization of both the probabilities $P_{X|Y}$ and $P_X$, \ie, one analyzes the entropies $\Huuy$ and $\Huu$ for general $P_{XY} = P_X P_{Y|X}$.
Note that $\Huuy \leq \Huu$.
It can be shown that the positions $i\in\bbN$ of $\vec u$ polarize into three sets \cite{Honda13}
\begin{align}
	\lim_{N\to\infty} \frac1N \abs{\mathcal H_{U|Y}}                     &= \He(X|Y) \\
	\lim_{N\to\infty} \frac1N \abs{\mathcal L_{U|Y} \cap \mathcal H_{U}} &= \MI(X; Y) \label{eq:hy_capacity} \\
	\lim_{N\to\infty} \frac1N \abs{\mathcal L_{U}}                       &= 1 - \He(X) \textnormal{,}
\end{align}
where $\mathcal H_U$ and $\mathcal L_U$ are defined similarly to $\mathcal H_{U|Y}$ and $\mathcal L_{U|Y}$, respectively, but where we drop the conditioning on $\vec Y$.
For $\Huu\approx 1$, the position $u_i$ can hold one bit of information given the previous bits $\vec u_\bbi$ and the distribution $P_X$.
For $\Huu\approx 0$, $u_i$ almost becomes a deterministic function of the previous bits in order to attain $P_X$ after polar-transforming the resulting $\vec u$.

To use this for coding, one partitions the indices into $\mathcal F$, $\mathcal I$ and $\mathcal D$ taking the role of $\mathcal H_{U|Y}$, $\mathcal L_{U|Y} \cap \mathcal H_U$, and $\mathcal L_U$, respectively.
The bits in $\mathcal F$ are again the ones with $\Huuy \approx 1$.
The ones in $\mathcal D$ are the positions with $\Huu \approx 0$.
The remaining positions in $\mathcal I$ have $\Huuy \approx 0$ and $\Huu \approx 1$.

By construction, the bits in $\mathcal F$ cannot be reliably recovered at the decoder and are frozen to $0$.
The probabilities $P_{U_i|\vec U_\bbi}(\cdot|\vec u_\bbi)$, $i\in\mathcal D$, are very close to either zero or one, making $u_i$ almost a deterministic function of $\vec u_\bbi$.
This allows the encoder to choose those bits in a deterministic and successive manner as \cite{Chou15,Mondelli18}
\begin{equation}
    u_i = \argmax_{u} P_{U_i|\vec U_\bbi}(u|\vec u_\bbi)
\end{equation}
which is efficiently performed by running an \gls{SC} decoder at the encoder.
Such an \gls{SC} encoder decodes the target distribution $P_X^N$ instead of the posterior probability $P_{X|Y}^N(\cdot|\vec y)$.
Finally, the bits in $\mathcal I$ can be arbitrarily chosen at the encoder and reliably decoded from $\vec y$ and are thus used to transmit the data.

This effectively constructs a generally non-linear subcode of the underlying linear polar code $\polarcode(\mathcal F)$.
The code consists of the code words $\vec x(\vec d)$ that the encoder outputs for each possible data vector $\vec d$.
The effective code is a proper subcode, \ie, there exist sequences $\vec x \in \polarcode(\mathcal F)$ which are not generated by the \gls{HY} encoder.

Analogously to decoding, the finite-length performance of the scheme can be improved by employing \gls{SCL}-based encoding with list size $L_\mathrm{enc}$.
In this case, the path metric is updated according to $P_{U_i|\vec U_\bbi}$ and the code word candidate $\vec x$ with highest $P_{\vec X}(\vec x)$ is chosen at the end.
As with the \gls{SC} encoder, the code word $\vec x$ is a deterministic function of the data and the target distribution $P_X$ also under \gls{SCL} encoding.
Note that the effective code book depends on the encoder and different list sizes may lead to different code books.

As \gls{HY} codes are subcodes of $\polarcode(\mathcal F)$, they can be decoded straightforwardly by using off-the-shelf decoders for linear polar codes.
For this, one sets $\mathcal R = \mathcal I \cup \mathcal D$ in \eqref{eq:sc_decoding}.
As both index sets have $\Huuy\approx 0$, the positions can be decoded correctly with high probability via \gls{SC} or \gls{SCL} decoding from the channel observation $\vec y$ and estimates of the previous bits $\vec{\hat u}_\bbi$.
After the bits $\vec{\hat u}_{\mathcal R}$ have been decoded, one discards the \gls{DM} bits $\vec{\hat u}_{\mathcal D}$ and outputs $\vec{\hat u}_{\mathcal I}$ as final decision on the data bits.
The above code construction as well as its polarization analysis extend readily to multilevel coding \cite{Runge22}.

\section{Proposed Decoder Modifications}
\label{sec:main}

This section describes a list decoder for \gls{HY} codes that exploits the deterministic nature of the encoder rather than merely discarding the bits $\vec{\hat u}_{\mathcal D}$.
We present two modifications, depending on whether the encoder uses \gls{SC} or \gls{SCL} procedures.
Both modifications improve the error rates by ensuring that the decoder estimate $\vec{\hat u}$ corresponds to a valid \gls{HY} code word $\vec{\hat x}$.

\subsection{Modified Decoding for SC Encoding}

\gls{SC} encoding chooses bit $u_i$, $i\in\mathcal D$, as a deterministic function of the previous bits $\vec u_\bbi$. The positions in $\mathcal D$ can thus be treated as non-linear dynamic frozen bits \cite{Trifonov13}, which means that all code word candidates of the modified decoder are valid code words of the \gls{HY} subcode.
An \gls{SC} decoder can readily check the validity of a partial candidate $\vec{\hat u}_{\idxset{i}}$ and correct invalid decisions.
Under list decoding, the decoder can reorder its candidate \gls{SC} paths using the dynamic frozen bit and continue decoding.

The implementation is as follows.
The standard \gls{SCL} decoder \cite{Tal15} has data structures to store soft information for each polarization stage and to store the already decided bits.
To compute the dynamic frozen bits, all data structures storing soft information are duplicated so that one has one set of structures for $P_{U_i|\vec U_\bbi}$ and another set for $P_{U_i|\vec U_\bbi, \vec Y}$.
The data structures used to recursively compute $P_{U_i|\vec U_\bbi, \vec Y}$ hold values computed from $P_{X|Y}^N(\cdot|\vec y)$.
Similarly, the data structures for $P_{U_i|\vec U_\bbi}$ hold values computed from $P_{X}^N$.
With this, the same probability calculations are performed on each set of data structures.
Similar to standard dynamic frozen bits, the \gls{DM} bits are then chosen deterministically as $\hat u_i = \argmax_{u} P_{U_i|\vec U_\bbi}(u|\vec{\hat u}_\bbi)$ and the path metric is updated based on $P_{U_i|\vec U_\bbi,\vec Y}(u_i|\vec{\hat u}_\bbi, \vec y)$.

The modified decoder does not alter the asymptotic computational complexity $\mathcal O(L_\mathrm{dec} \cdot N \log N)$ of the \gls{SCL} decoder with list size $L_\mathrm{dec}$.
Although the number of soft information calculations is doubled, the list sorting step, which usually dominates the computational effort, is left untouched.

\subsection{Modified Decoding for SCL Encoding}

The bits $u_i$, $i\in\mathcal D$, may also depend on ``future'' bits if the transmitter employs list encoding.
These bits can thus not be treated as dynamic frozen bits.
Instead, we remove invalid candidates in a final step after completing standard \gls{SCL} decoding and after checking the \gls{CRC}. To check if a list candidate is a valid code word of the \gls{HY} subcode, we extract the estimated $\vec{\hat u}_{\mathcal I}$ and re-encode these with the same encoder as the one at the transmitter.
If the re-encoded code word differs from the decoded candidate, the candidate is removed from the list.
The candidate code words can be checked in parallel.
Re-encoding one candidate with list size $L_\mathrm{enc}$ has a computational complexity of $\mathcal O(L_\mathrm{enc} \cdot N \log N)$.
The computational complexity of parallel validity checking under \gls{SCL} decoding with list size $L_\mathrm{dec}$ is thus $\mathcal O((L_\mathrm{dec} + L_\mathrm{enc}L_\mathrm{dec}) \cdot N \log N)$.

To ease memory requirements, the re-encoding can be performed sequentially.
For this, the list of code word candidates is sorted by path metric.
The decoder checks the candidates in order from most likely to least likely, chooses the first valid one, and stops decoding.
The \gls{SCL} decoding logic and large parts of the same memory can be reused for re-encoding.
This results in the following algorithm.
\begin{enumerate}
    \item Perform \gls{SCL} decoding of $\vec y$ yielding a list of $L_\mathrm{dec}$ candidates $\vec{\hat x}^{(\ell)} = \vec{\hat u}^{(\ell)} \matr G_N$, $\ell\in\idxset{L_\mathrm{dec}}$.
    \item For each $\ell \in \idxset{L_\mathrm{dec}}$ sorted by path metric, \begin{enumerate}
        \item check \gls{CRC} for validity and proceed to next candidate if invalid,
        \item extract data estimate $\vec{\hat u}^{(\ell)}_{\mathcal I}$ and re-encode to $\vec x(\vec{\hat u}^{(\ell)}_{\mathcal I})$,
        \item if $\vec{\hat x}^{(\ell)}$ and $\vec x(\vec{\hat u}^{(\ell)}_{\mathcal I})$ are equal, return candidate $\ell$ as decoder estimate.
    \end{enumerate}
\end{enumerate}

Compared to parallel re-encoding the computational overhead is improved for two reasons:
First, re-encoding more than one candidate is only necessary if the unmodified \gls{SCL} decoding would output an incorrect decision.
This is because the transmitted code word is always a valid HY code word.
The average number of re-encodings therefore reduces with $\mathrm{FER}_\mathrm{SCL}$.
Second, %
even if the first candidate is not valid, the transmitted code word will with high probability be among the most likely candidates so that only very rarely all candidates need to be re-encoded.
The average number of $\mathcal O(N\log N)$-complexity \gls{SC} passes is $L_\mathrm{dec}$ for \gls{SCL} decoding plus
\begin{equation}
    L_\mathrm{enc}\cdot (1-\mathrm{FER}_\mathrm{SCL}) + \E[\Lambda] \cdot L_\mathrm{enc} \cdot \mathrm{FER}_\mathrm{SCL}
    \label{eq:compl}
\end{equation}
for re-encoding the candidates, where the random variable $\Lambda$ denotes the
number of re-encodings conditioned on the event that the \gls{SCL} decoder fails.
Thus, for small \glspl{FER} the modification incurs almost no overhead in asymptotic growth of complexity.
Furthermore, numerical simulations will show that $\Lambda$ is significantly smaller than $L_\mathrm{dec}$ in many cases.

\section{Numerical Results}
\label{sec:results}
\subsection{System Model}

To evaluate the performance of our decoder, we perform \gls{MC} simulations of an end-to-end \gls{MLHY} system.
We now denote with $\vec X$ a code word of the \gls{MLHY} code, \ie, a modulated code word.
We consider binary uniform \gls{iid} input data and encode it to words of \gls{ASK} or \gls{PAM} symbols.
The corresponding modulation alphabets of cardinality $M = 8$ and $M=4$ are
\begin{align*}
	\mathcal{X}_\text{8-ASK} = \{\pm 7, \pm 5, \pm 3, \pm 1\}, \quad
	\mathcal{X}_\text{4-PAM} = \{0,1,2,3\} \textnormal{,}
\end{align*}
respectively.
For both cases, we choose $P_X(x) \propto \exp(-\nu\abs{x}^2)$ with parameter $\nu$ and use set-partitioning labelling \cite{Wachsmann99}.
The code words are transmitted over a real-valued \gls{AWGN} channel
\begin{equation}
    Y = X + N
\end{equation}
with zero-mean Gaussian noise $N$.
We define the \gls{SNR} as $\frac{\E[\abs{X}^2]}{\E[\abs{N}^2]}$.
The \gls{SNR} and the effective channel input distribution $P_X$ are known at the decoder.

\subsection{Code Construction}

We construct codes as described in \cite[Sec.~IV.]{Runge22}.
Given the design parameters $K$ and $N_\mathrm{DM}$, we choose the $N_\mathrm{DM}$ positions with smallest $\Huu$ as elements of $\mathcal D$.
The positions in $\mathcal I$ are then the $K$ channels with smallest $\Huuy$ among the remaining indices.
The bits $\vec u_{\mathcal I}$ may also include a \gls{CRC} checksum of length $p$, which can be optimized by exhaustive search.
All other indices become elements of $\mathcal F$.
The resulting code has rate $R = \frac{K - p}{N}$.

As in \cite{Runge22}, we introduce two more design parameters to specify the entropies $\Huu$ and $\Huuy$.
The first is the \gls{dSNR} that determines the noise variance for estimating $\Huuy$.
The second is a parameter $\kappa$ for code optimization and to choose the rate-optimal $P_X^*$ at $\kappa\cdot\mathrm{dSNR}$ as the channel input distribution for which we estimate $\Huu$.
Both parameters can improve the performance for finite length shaping.

The entropies are easily estimated via \gls{MC} simulation for block lengths up to several thousand bits as follows.
First, $\Huu$ is estimated for $P_X^*$ using a \gls{SC} decoder \cite{Runge22}.
After choosing the $N_\mathrm{DM}$ positions for $\mathcal D$, the effective channel input distribution $P_X$ is measured and then used to estimate the reliabilities $\Huuy$ from samples of an \gls{AWGN} channel with the \gls{dSNR}.
Instead of using \gls{MC} integration, one can also estimate the entropies via density evolution \cite{Trifonov22} or a \gls{BEC} approximation \cite{Wu23}.

\subsection{Frame Error Rates}

\begin{figure}[tb]
	\centerline{\includegraphics{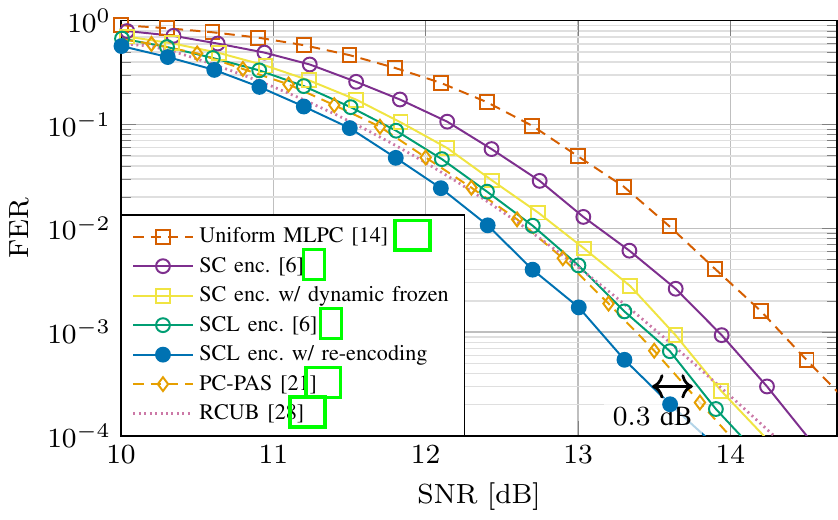}}
	\caption{Performance of MLHY coding under modified and standard decoding compared to PC-PAS, uniform MLPC, and the RCUB, with a $4$-PAM constellation, $N=64$, at a $R = \SI{1.25}{\bpcu}$, shaped $\mathrm{dSNR} = \SI{18.1}{\dB}$, $\kappa= \SI{-0.9}{\dB}$, $N_\mathrm{DM} = 24$, and uniform $\mathrm{dSNR} = \SI{19.25}{\dB}$. The MLHY and MLPC codes do not use an outer CRC, PC-PAS a 3-bit CRC. For PC-PAS, $\mathrm{dSNR} = \SI{14.5}{dB}$ and $\kappa = \SI{-3.9}{dB}$.}
	\label{fig:fer_pam4}
\end{figure}

We consider \gls{SCL} decoding with $L_\mathrm{dec} = 32$ together with \gls{SC} encoding or \gls{SCL} encoding with $L_\mathrm{enc} = 32$.
Fig.~\ref{fig:fer_pam4} shows the \glspl{FER} of \gls{MLHY} coding under various encoder and decoder configurations for $4$-\gls{PAM}.
First, we focus on the proposed dynamic frozen bit modified decoder for \gls{SC} encoding~\pltref{plt:fer_pam4_m1}.
The error performance improves over standard \gls{SCL} decoding~\pltref{plt:fer_pam4_basic} and almost approaches the performance of \gls{SCL} encoding with a standard decoder~\pltref{plt:fer_pam4_mlhy} \cite{Runge22}.
Thus, the modified decoding can be used to shift computational complexity from the transmitter to the receiver in system design.
Employing \gls{SCL} encoding at the transmitter and re-encoding at the receiver~\pltref{plt:fer_pam4_m32} further improves the \gls{FER}.
For both, \gls{SC} and \gls{SCL} encoding, the additional validity checking yields gains of approximately \SI{0.3}{\dB} over standard \gls{SCL} decoding.
Furthermore, candidate re-encoding allows \gls{MLHY} coding to perform better than \gls{PCPAS} with the \gls{CCDM} type-check.
Our system surpasses the \gls{RCUB}~\pltref{plt:fer_pam4_rcub} \cite{Polyanskiy10} by \SI{0.45}{\dB}.

\tikzsetnextfilename{fer_pam4}

\begin{figure}[tb]
	\centerline{\includegraphics{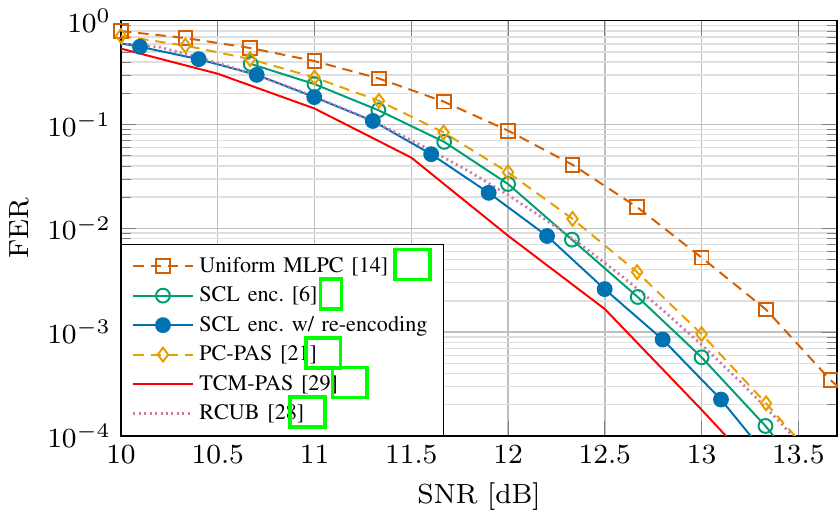}}
	\caption{Performance of MLHY coding under modified and standard decoding compared to PC-PAS, TCM-PAS, uniform MLPC, and the RCUB, with an $8$-ASK constellation, $N=64$, at a $R = \SI{1.75}{\bpcu}$, shaped $\mathrm{dSNR} = \SI{13}{\dB}$, $\kappa = \SI{-1}{\dB}$, $N_\mathrm{DM} = 23$, and uniform $\mathrm{dSNR} = \SI{13}{\dB}$. The MLHY and MLPC codes use an outer 7-bit CRC, the codes for modified decoding a 5-bit CRC, and PC-PAS a 4-bit CRC together with a type check. For PC-PAS, $\mathrm{dSNR}=\SI{8}{dB}$ and $\kappa=\SI{-0.6}{dB}$.}
	\label{fig:fer_ask8}
\end{figure}

Fig.~\ref{fig:fer_ask8} depicts the performance for $8$-ASK.
In this scenario, the gains from checking code word candidate validity are smaller.
Re-encoding at the receiver~\pltref{plt:fer_pam4_m32} still gains approximately \SI{0.1}{\dB} over standard \gls{SCL} decoding.
The gain for \gls{SC} encoding and checking via dynamic frozen bits is slightly smaller.
We also compare our algorithm to \gls{TCMPAS}~\pltref{plt:fer_ask8_tcm} \cite{Wang23}, which achieves better performance for short block lengths at the expense of larger but adaptive list sizes.
While their scheme performs even further beyond the \gls{RCUB}, it is restricted to symmetric constellations due to the \gls{PAS} architecture.

\tikzsetnextfilename{fer_ask8}

The decoding gains from checking the validity of code word candidates seem to decrease with increasing block length. For large block lengths, the bit channel polarization is stronger and there are fewer weakly polarized bit channels that introduce erroneous paths at a standard \gls{SCL} decoder.

\subsection{Computational Overhead}

\begin{figure}[tb]
	\centerline{\includegraphics{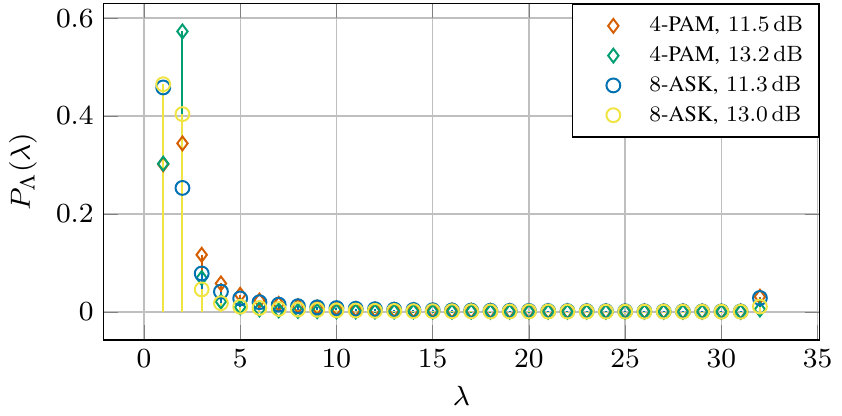}}
	\caption{Empirical distribution of $\Lambda$ at different SNR.}
	\label{fig:mlhy_reenc_compl}
\end{figure}

We analyze the computational overhead for received sequences for which standard \gls{SCL} decoding would fail, \ie, the second term in \eqref{eq:compl}.
A worst case analysis showed that it is possible that all but the last $L_\mathrm{dec}-1$ candidates have to be re-encoded.
Fig.~\ref{fig:mlhy_reenc_compl} shows the empirical distribution of $\Lambda$ for cases discussed in the previous section.
We observe that on average, the number of re-encodings is smaller than its maximum $L_\mathrm{dec}$.
In fact, here most of the probability mass concentrates around $\Lambda < 5$.
From our simulations, the peak at $\Lambda = 32$ corresponds to the cases where there was no valid candidate code word in the list.

\tikzsetnextfilename{reenc_compl}

\section{Conclusions}

\gls{SCL} decoder modifications for polar-coded shaping were presented that provide coding gains of up to \SI{0.3}{\dB} at short block lengths with negligible penalty in complexity.
The modifications work with any \gls{HY}-based code.

\section*{Acknowledgment}
The authors wish to thank Gerhard Kramer for suggestions.

\bibliographystyle{IEEEtran}
\bibliography{IEEEabrv,confs-jrnls,manual}

\end{document}